# Spatiotemporal coupled Airy-Airy wavepacket and its propagation dynamics

ZHAOFENG HUANG,[1,†] XIAOLIN SU,[1,†] QIAN CAO,[1,2,*] ANDY CHONG,[3,4] AND QIWEN ZHAN[1,2,5]

[1]*School of Optical-Electrical and Computer Engineering, University of Shanghai for Science and Technology, Shanghai 200093, China*
[2]*Zhangjiang Laboratory, Shanghai 201210, China*
[3]*Department of Physics, Pusan National University, Busan 46241, Republic of Korea*
[4]*Institute for Future Earth, Pusan National University, Busan 46241, Republic of Korea*
[5]*Westlake Institute for Optoelectronics, Fuyang, Hangzhou, China*
[†]*These authors contributed equally.*
*\*cao.qian@usst.edu.cn*

**Abstract:** Airy beams, celebrated for their self-acceleration, diffraction-free propagation, and self-healing properties, have garnered significant interest in optics and photonics, with applications spanning ultrafast optics, laser processing, nonlinear optics, and optical communications. Recent research primarily aims at independent control of Airy beams in both spatial and spatiotemporal domains. In a pioneering approach, we have successfully generated and controlled a spatiotemporal coupled (STc) Airy-Airy wavepacket, achieving its rotation while preserving vertical distribution in the spatiotemporal domain. Furthermore, we have investigated the self-acceleration and self-healing properties of the STc Airy-Airy wavepacket in this domain, noting that its dynamically adjustable rotation and spatiotemporal coupling capability provide a novel strategy for managing ultrafast lasers, with potential advancements in optical micromanipulation and time-domain coding communication.

## 1. Introduction

Diffraction and dispersion, as fundamental phenomena in light wave propagation, cause the spatial and temporal broadening of wave packets, which have always been the core challenges in beam manipulation, information transmission, and the improvement of imaging resolution. Airy beams [1] have been proven to be the only non-dispersing, localized solutions in one dimension, and Airy beams with infinite energy can achieve diffraction-free propagation [2]. The physical properties of Airy beams originate from their asymmetric Airy function distribution and the cubic phase modulation in the Fourier space. In addition to the diffraction-free property, Airy beams have other interesting properties, such as self-acceleration [3,4] and self-healing [5]. Although the intensity centroid of these beams always moves along a straight line, the actual characteristic of the intensity structure of these wavefronts is indeed transverse acceleration. Moreover, their self-healing ability enables them to reconstruct the waveform after being blocked or scattered. The optical Airy beams with finite energy were first generated in the laboratory in 2007 [6] by loading a cubic phase in the Fourier plane of a Gaussian beam through a liquid-crystal spatial light modulator. This has promoted extensive research on Airy beams after [7]. Airy beams show numerous applications in beam shaping [8], particle manipulation [9–11], plasma generation [12,13], biomedical imaging [14], ultrafast optics [15], and laser machining [16], as well as optical communications [17,18].

In recent years, the study of spatiotemporal optical fields has garnered significant attention due to its introduction of a coupled spatial and temporal distribution for the optical field [19–21]. This enables the optical field to achieve distinct manipulations and exhibit special propagation dynamics that were not previously attainable. However, the majority of research on Airy beams has predominantly focused on their spatial propagation dynamics [22,23]. Moreover, even in studies involving spatiotemporal Airy-Airy wavepackets [24,25], the temporal and spatial aspects are often treated as independent variables [26]. The study on the spatiotemporal coupled Airy-Airy wavepacket remains unexplored.

In this work, we generated and observed the STc Airy-Airy wavepacket in the laboratory, for the first time. We studied the propagation dynamics of STc Airy-Airy wavepacket, including its spatiotemporal self-acceleration and self-healing properties. The spatiotemporal coupling enables a precise control over the dynamic trajectory of the spatiotemporal light field. The results demonstrate that the STc Airy-Airy wavepacket is immune against diffraction, dispersion, and perturbations. The wavepacket can maintain its spatiotemporal profile over extended propagation distances of 20 cm, which corresponds to 2x diffraction length and dispersion length. These features present new opportunities for this wavepacket in applications in obstacle avoidance, interference suppression, and precise guidance, with considerable potential for advancements in high-speed information transmission, particle manipulation, and laser processing [27].

## 2. Methods and Results

The one-dimensional Airy function, characterized by its infinite oscillatory tail, serves as an accelerating solution to the (1+1)D free-space Schrödinger equation [4]. The mathematical resemblance between paraxial diffraction in space and second-order dispersion in time allows for the generation and study of spatiotemporal optical Airy waves in dispersive systems [28]. Notably, under normal dispersion conditions, the sign of the time operator is opposite to that of the spatial operator [29]. In a (2+1)D dispersive optical paraxial system, the dispersive effects can be associated with the term involving the moving time coordinate $\tau = t - \frac{z}{v_g}$, the wavenumber is $k = \frac{2\pi n}{\lambda}$, and the second-order dispersion coefficient at $\omega_0$ is $k'' = \frac{d^2k}{d\omega^2}\big|_{\omega=\omega_0}$. For a conventional uncoupled spatiotemporal Airy wavepacket, its field distribution is the product of a temporal Airy pulse and a spatial Airy beam, which can be represented by the following expression [30]

$$\psi(X,T,z) = Ai\left(X - \frac{\alpha^2 z^2}{4}\right) Ai\left(T - \frac{\beta^2 z^2}{4}\right) \exp\left[\frac{iz}{2}(\alpha X - \beta T) + \frac{iz^3}{12}(\beta^3 - \alpha^3)\right], \quad (1)$$

where $\alpha = \frac{1}{k(\Delta x)^2}$, $\beta = \frac{k''}{(\Delta \tau)^2}$. The normalized coordinates $(X,T)$ are related to the space and time coordinate $(x,t)$ via $T = \frac{t}{\Delta \tau}$, $X = \frac{x}{\Delta x}$. $\Delta x$ and $\Delta \tau$ are the width of the wavepacket in the spatial domain and in the temporal domain. Equation (1) shows the expression of spatiotemporal un-coupled Airy-airy wavepacket. When the Airy function is applied in a coupled way between the spatial coordinate and the temporal coordinate, the resulting wavepacket will have different propagation dynamics due to the joint effect of dispersion and diffraction.

The generation of spatiotemporal coupled Airy-Airy wavepacket can be accomplished by applying a spatiotemporally correlated cubic phase in the Fourier plane $(\omega, k_x)$ of the wavepacket. The propagation function of the spatiotemporal coupled Airy-Airy wavepacket can be expressed as

$$\psi'(X,T,z) = Ai\left[\xi(X,T) - \frac{\alpha^2 z^2}{4}\right] Ai\left[\eta(X,T) - \frac{\beta^2 z^2}{4}\right]$$

$$\times \exp\left(\frac{iz}{2}[\alpha\xi(X,T) - \beta\eta(X,T)] + \frac{iz^3}{12}(\beta^3 - \alpha^3)\right), \quad (2)$$

Here, $\xi(X,T)$ and $\eta(X,T)$ are given by

$$\begin{bmatrix} \xi(X,T) \\ \eta(X,T) \end{bmatrix} = \begin{bmatrix} \cos\theta & \sin\theta \\ -\sin\theta & \cos\theta \end{bmatrix} \begin{bmatrix} X \\ T \end{bmatrix}, \quad (3)$$

In this equation, $\theta$ represents the rotation angle of the spatiotemporal Airy beam in the $x - t$ plane.

STc Airy-Airy wavepacket can maintain its self-acceleration dynamics in the spatial domain and in the temporal domain. It is noteworthy that the self-acceleration in both spatial and temporal dimensions is now influenced by the rotation angle $\theta$. Combining Eqn. (2) and Eqn. (3), it can be derived that the main lobe of the STc Airy-Airy wavepacket follows the trajectories satisfying the following self-accelerating curve equations,

$$X = \frac{(\alpha^2 \cos\theta - \beta^2 \sin\theta) \cdot Z^2}{4} + X_0, \quad (4)$$

$$T = \frac{(\alpha^2 \sin\theta + \beta^2 \cos\theta) \cdot Z^2}{4} + T_0. \quad (5)$$

Here, $T_0$ and $X_0$ are the initial positions for the main lobe of the wavepacket. In addition, the STc Airy-Airy wavepacket maintains its non-spreading feature during dispersive propagation, resisting spatial and temporal broadening. The key advantage of the rotated Airy-Airy wavepacket over the traditional Airy-Airy wavepacket is its self-healing ability. Rotating the STc Airy-Airy wavepacket at an arbitrary angle can enable the side lobes to better avoid obstacles or minimize the blockage, thus enhancing the beam's self-healing efficiency.

In the experiment, we consider a finite-power Airy function with an exponential decay factor $a$ ($a > 0$), whose wave packet exhibits a Gaussian power spectrum because its Fourier transform in the normalized k-space can be written as

$$\psi(k) = \exp(-ak^2) \cdot \exp[i/3(k^3 - 3a^2k - ia^3)]. \quad (6)$$

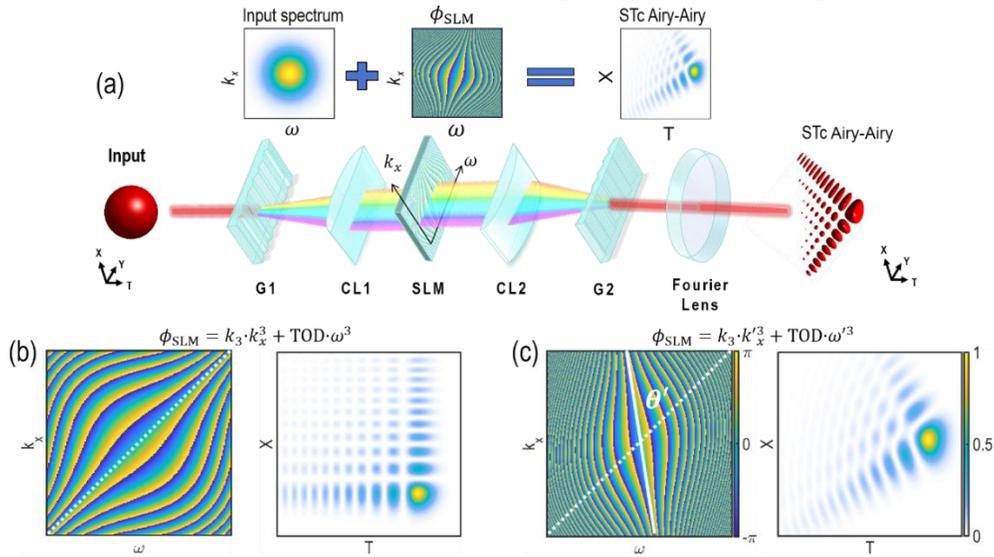

**Figure 1 Generation of spatiotemporal coupled (STc) Airy-Airy wavepacket.** (a) Fourier phase loaded on spatial light modulator (SLM) $\phi_{SLM}(\omega, k_x)$ for generating STc Airy-Airy wavepacket. Schematic for generating spatiotemporal coupled Airy-Airy wavepacket. The phase is loaded in the Fourier plane ($\omega - k_x$ plane) of the input light field in a spatiotemporal

pulse shaper setup[19]. (b) Fourier phase and the intensity distribution for a spatiotemporal un-coupled Airy-Airy wavepacket. (c) Fourier phase and the intensity distribution for an STc Airy-Airy wavepacket. When the Fourier phase is rotated by $\theta'$ in the normalized $\omega - k_x$ plane.

In the experiment, we managed to generate and study the STc Airy-Airy wavepacket. Figure 1 illustrates the generation process and results. Using a diffraction grating and a cylindrical lens, we performed a one-dimensional temporal Fourier transform on the initial Gaussian beam, allowing the SLM plane to function as the $\omega - k_x$ plane. Then we applied first-order and third-order phases $\phi_{SLM}(\omega, k_x)$ to the SLM, as depicted in Fig. 1(a). Next, an inverse temporal Fourier transform was executed through the spatiotemporal pulse shaper. After the pulse shaper, a Fourier lens conducted a spatial Fourier transform on the modulated beam, producing a spatiotemporal Airy-Airy wavepacket at the back focal plane of the lens. The phase applied to the SLM and the shape of spatiotemporal un-coupled Airy-Airy wavepacket is shown in Fig. 1(b). In the figure, $k_3$ and $TOD$ are cubic (third-order) phase coefficients for $k_x$ and $\omega$, respectively.

To generate a spatiotemporal coupled Airy-Airy wavepacket with a rotation angle $\theta$, we applied first-order and third-order phases with spatiotemporal coupling rotational degree of freedom to the SLM. The phase applied to the SLM is represented by

$$\phi'_{SLM} = k_3 k_x^{'3} + TOD \omega^{'3}, \tag{7}$$

where $k_x'$ and $\omega'$ are given by

$$\begin{bmatrix} k_x' \\ \omega' \end{bmatrix} = \begin{bmatrix} \cos\theta & \sin\theta \\ -\sin\theta & \cos\theta \end{bmatrix} \begin{bmatrix} k_x \\ \omega \end{bmatrix}. \tag{8}$$

When $\theta = \frac{\pi}{6}$, the generated rotated STc Airy-Airy wavepacket is shown in Fig. 1(c). The Fourier phase is rotated by $\theta'$ in a normalized $(\omega', k_x')$ plane, resulting in spatiotemporal Airy-Airy wavepacket is rotated by an angle of $\theta$ in the normalized spatiotemporal domain. The relationship between two angle is given by

$$\theta' = \tan^{-1}\left(\frac{\sin\theta - \sqrt[3]{\frac{TOD}{k_3}}\cos\theta}{\cos\theta + \sqrt[3]{\frac{TOD}{k_3}}\sin\theta}\right) - \tan^{-1}\left(-\sqrt[3]{\frac{TOD}{k_3}}\right). \tag{9}$$

### 3. Experimental Results

The experimental setup is depicted in Fig. 2 for generating STc Airy-Airy wavepacket and studying its propagation dynamics. A home-built fiber laser emits a pulsed Gaussian beam that is divided into the object beam and the probe beam by a beam splitter (BS1). In the "object" arm (blue beam path in Fig. 1), the pulse enters a spatiotemporal pulse shaper consisting of a diffraction grating (G3), a cylindrical lens, and a liquid crystal spatial light modulator (SLM). After the spatiotemporal pulse shaper, the output pulse is focused by a Fourier lens to the CCD camera. The Fourier lens has a focal length of 400 mm and the CCD is placed at the back focal plane of the Fourier lens. In such a configuration, the SLM can modulate the Fourier phase $\phi_{SLM}(\omega, k_x)$ of the wavepacket.

In the other arm, the "probe" arm, the pulse is compressed by a pair of gratings (G1 and G2) to the transform-limited form. The probe pulse is then sent to the CCD camera with a controllable delay line control. At the CCD, the "object" light and the "probe" pulse are re-combined spatiotemporally. Scanning the relative delay between two pulses can produce a series of fringe images at the CCD. Using the same algorithm used in [31], we can retrieve the three-dimensional profiles of the generated STc Airy-Airy wavepacket.

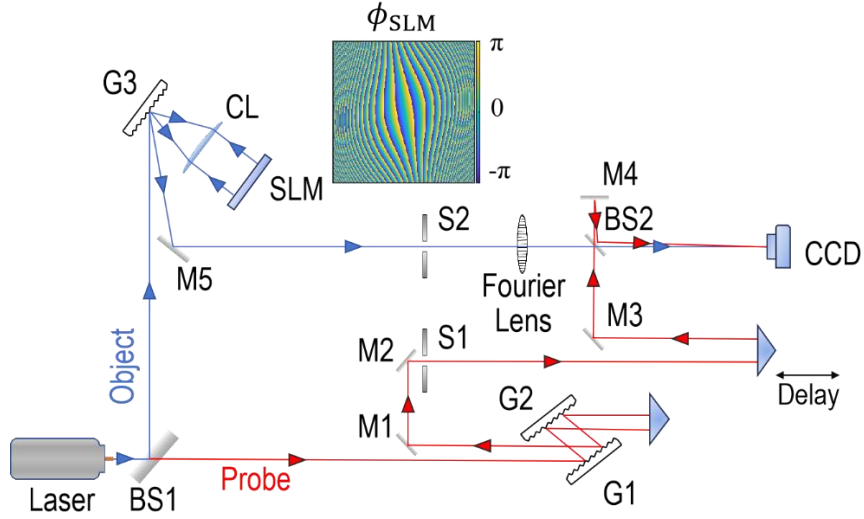

**Figure 2 Experimental setup for observing and studying STc Airy-Airy wavepacket.** The setup has a Mach-Zehnder interferometric structure. The laser light is split into the "probe" arm (red beam path) and the "object" arm (blue beam path) by BS1. The "probe" light is compressed by a grating pair pulse compressor (G1 and G2) and is delivered by a controllable delay line stage. The "object" light is modulated by a spatiotemporal pulse shaper (G1, CL, and SLM) and is focused by a Fourier lens. In such a configuration, the angular-spectral phase of the light field $\phi(\omega, k_x)$ is modulated by the SLM. The "object" and the "probe" light are recombined after BS2, spatiotemporally, at the back focal plane of the Fourier lens, where the CCD camera is placed. The interferometric information between them can be used to reconstruct the generated STc Airy-Airy wavepacket[31]. M: mirror; BS: beamsplitter; G: grating; CL: cylindrical lens; SLM: spatial light modulator; S: shutter; wavepacket.

Using this setup, we first generate the STc Airy-Airy wavepacket at the back focal plane of the Fourier lens, where we define this position as $z = 0 \text{ mm}$. The leftmost figure in Fig. 3(a) shows the generated STc Airy-Airy wavepacket. The characteristic multi-lobe Airy structure is observed in the spatiotemporal domain in a coupled fashion. To study the propagation dynamics of the wavepacket, we propagate it in a "virtual" dispersive medium with a dispersion coefficient of $1.6 \times 10^3 \text{ fs}^2/\text{mm}$. To reference the wavepacket to a spatiotemporal origin, we also propagate an un-modulated Gaussian-Gaussian wavepacket with the same propagation setting. The center of mass position of the un-modulated Gaussian-Gaussian wavepacket is used for calibrating the spatiotemporal origin. Figure 3(a) shows the spatiotemporal self-acceleration of the STc Airy-Airy wavepacket for a total propagation distance of 200 mm. Arrows are added to the plots to illustrate the direction of self-acceleration. In a total propagation distance of 200 mm, the central lobe has shifted 100 μm in the X-direction and 490 fs in the T-direction.

To quantitatively assess its acceleration behavior, we measured the change of the wavepacket in an interval of 25 mm in the propagation distance, focusing on its main lobe position. The results for its main lobe shift in the spatial domain and in the temporal domain are presented in Fig. 3(b) and Fig. 3(c). To compare, the numerical simulation results (2D colored profile) and the theoretically calculated results (blue solid lines) are also added to the figures. These results agree with each other well. This also indicates a new opportunity for enhanced control over the propagation of Airy beams, offering potential applications in particle manipulation, beam obstacle avoidance, and other areas.

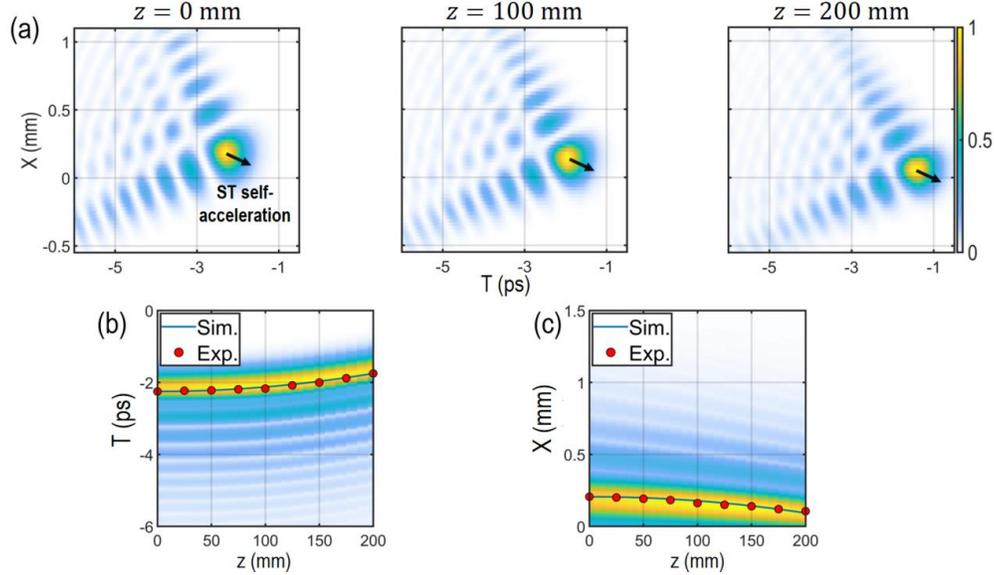

**Figure 3 Spatiotemporal self-acceleration of the STc Airy-Airy wavepacket.** At the back focal plane of the Fourier lens, where $z = 0$ mm is defined, we observe the formation of STc Airy-Airy wavepacket. Further propagating the wavepacket can result in the spatiotemporal self-acceleration of the wavepacket. (a) Measured spatiotemporal intensity profile of the wavepacket at $z = 0$ mm, 100 mm and 200 mm. The origin of the spatiotemporal plane is defined by measuring the central position of an un-modulated Gaussian-Gaussian wavepacket in the same propagation setting. (b) Self-acceleration of the main lobe wave in the temporal direction. (c) Self-acceleration of the main lobe wave in the spatial direction.

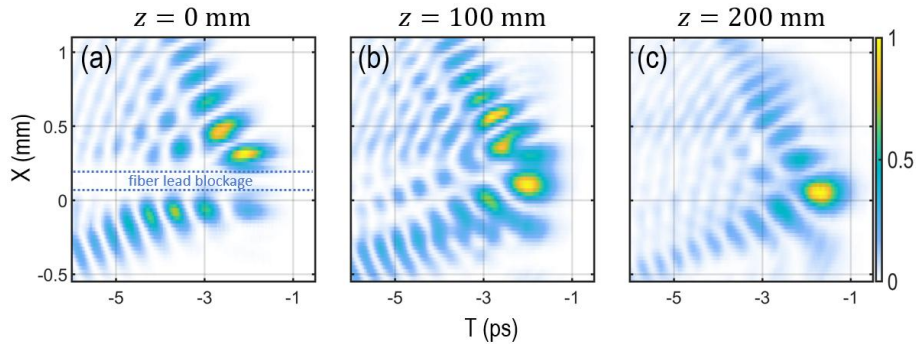

**Figure 4 Self-healing of the spatiotemporal Airy-Airy wavepacket.** To observe the spatiotemporal self-healing of the STc Airy-Airy wavepacket, we placed a fiber lead with a diameter of 125 µm slightly before the Fourier plane (back focal plane of the Fourier lens, where $z = 0$ mm is defined) so that the main lobe of the wavepacket is blocked. After a dispersive propagation of $z = 200$ mm, the measured spatiotemporal profile reveals the spatiotemporal self-healing process of the wavepacket.

Apart from self-acceleration, we also investigated the self-healing properties of the spatiotemporal coupled Airy-Airy wavepacket, which is another intriguing feature of the optical Airy wave. To implement, we put an optical fiber lead slightly before the back focal plane of the Fourier lens for partially blocking the wavepacket. The blocked wavepacket is measured to have a profile that is shown in Fig. 4(a). Further propagating the wavepacket can lead to its self-healing process. The wavepacket is measured at $z = 100$ mm and $z = 200$ mm. The results, as shown in Fig. 4, reveal a progressive recovery of the main lobe and a degree of self-healing during the propagation process. Moreover, Fig. 4 demonstrates that even with the main lobe obstructed, the STc Airy-Airy wavepacket continues to exhibit self-acceleration in both spatial and temporal directions during the self-healing process.

The primary distinction from prior studies is the spatiotemporal coupling of the Airy wavepacket. This coupling influences the beam's propagation trajectory through the interaction between time and space, allowing for more intricate trajectory control. For instance, by incorporating rotation or dynamically altering the STc Airy-Airy wavepacket's direction, the trajectory can be adjusted in real-time to adapt to varying spatiotemporal conditions, demonstrating enhanced adaptability and flexibility. Conversely, the traditional Airy-Airy beam, despite its ability to circumvent obstacles, follows a fixed trajectory.

Our control method introduces additional spatiotemporal coupling rotational degree of freedom to the conventional time and space dimensions, enhancing light field control and integration. Consequently, the STc Airy-Airy beam can convey more information than the traditional Airy beam, offering increased potential for applications in optical communication. By integrating the spatiotemporal joint rotation effect, the Airy beam we examined demonstrates enhanced self-healing capabilities and improved propagation dynamics in complex environments. We first noted that it can dynamically circumvent obstacles while preserving propagation quality. This suggests that the beam's imaging and transmission are more efficient and possess superior anti-interference properties. The STc Airy-Airy beam could offer substantial benefits in laser imaging and optical transmission.

## 4. Conclusions

We have experimentally generated the spatiotemporal coupled Airy-Airy wavepacket, which demonstrates distinct propagation dynamics due to the spatiotemporal coupling between the Airy functions. Compared to traditional spatiotemporal un-coupled instance, its energy envelope follows a self-accelerating trajectory through joint spatiotemporal regulation and can revert to its initial mode after disturbances, exhibiting structural stability and self-healing.

This coupling mechanism grants the beam multi-dimensional dynamics, enabling precise control of its trajectory in both time and space, and supporting arbitrary planar rotational motion. These features significantly enhance its interference resistance and manipulation flexibility in complex environments. Consequently, the spatiotemporally coupled Airy beam holds significant application potential in ultrafast laser processing, optical communication, and microscopic imaging, establishing an experimental foundation for developing new spatiotemporal coupled optical field systems.

**Acknowledgment**. We acknowledge financial support from National Natural Science Foundation of China (NSFC) [Grant Nos. 12434012 (Q.Z.) and 12474336 (Q.C.)], the Shanghai Science and Technology Committee [Grant Nos. 24JD1402600 (Q.Z.) and 24QA2705800 (Q.C.)], National Research Foundation of Korea (NRF) funded by the Korea government (MSIT) [Grant No. 2022R1A2C1091890], and Global - Learning & Academic research institution for Master's·PhD students, and Postdocs (LAMP) Program of the National Research Foundation of Korea(NRF) grant funded by the Ministry of Education [No. RS-2023-00301938]. Q.Z. also acknowledges support by the Key Project of Westlake Institute for Optoelectronics [Grant No. 2023GD007].

**Data availability.** Data underlying the results presented in this paper are not publicly available at this time but may e obtained from the authors upon reasonable request.

**Disclosures.** The authors declare no conflicts of interest.